\documentclass[11pt]{aastex}
\usepackage{graphicx,emulateapj5,apjfonts}
\usepackage{onecolfloat}
\usepackage{epsf}
\usepackage[hyperindex,breaklinks]{hyperref}

\shortauthors{Bell}
\shorttitle{Galaxy bulges and quenching}

\slugcomment{{\sc The Astrophysical Journal: } to appear August 1, 2008 }

\begin{document}

%%%%% Added the \def\head{ lark.

\def\head{

\title{Galaxy bulges and their black holes: a requirement
for the quenching of star formation}

\author{Eric F.\ Bell}
\affil{Max-Planck-Institut f\"ur Astronomie,
K\"onigstuhl 17, D-69117 Heidelberg, Germany; \texttt{bell@mpia.de}}

\begin{abstract}

One of the central features of the last 8 to 10 billion years of 
cosmic history has been the emergence of a well-populated 
red sequence of non-star-forming galaxies.  A number of models of galaxy 
formation and evolution have been devised to attempt to explain
this behavior.  Most current models require 
feedback from supermassive black holes (AGN feedback) to 
quench star formation in galaxies in the centers of their
dark matter halos (central galaxies).  
Such models make 
the strong prediction that 
all quenched central galaxies must have
a large supermassive black hole (and, by association, a prominent bulge 
component).  I show using data from the   
Sloan Digital Sky Survey that the observations are consistent 
with this prediction.  
Over 99.5\% of red sequence galaxies with 
stellar masses in excess of $10^{10} M_{\sun}$ have a prominent 
bulge component (as defined by having a S\'ersic index $n$ above 
1.5).  Those very rare red sequence central 
galaxies with little or no bulge ($n<1.5$) usually
have detectable star formation or AGN activity; the fraction of 
truly quenched bulgeless central galaxies is 
$<0.1\%$ of the total red sequence population.
I conclude that a bulge, and by implication a supermassive black
hole, is an absolute requirement for full quenching of 
star formation in central galaxies.  This is in agreement with the 
most basic prediction of the AGN feedback paradigm.
\end{abstract}

\keywords{galaxies: evolution --- galaxies: bulges --- 
galaxies: spiral --- galaxies: general ---  
galaxies: stellar content }
}%%%end head

\twocolumn[\head]

\section{Introduction}

A key feature of the last 8--10 billion years of galaxy 
evolution is the emergence of a well-populated red sequence of 
non-star forming galaxies \citep[e.g.,][]{bell04c17,faber06,brown07}.
The present-day red sequence has little scatter in color at 
a given magnitude, and appears to be largely composed of bulge-dominated
and early-type spiral galaxies \citep{vis,ble2}.  
Theoretical models of galaxy evolution
have had difficulty in reproducing the properties of red sequence 
galaxies using standard recipes for the cooling of 
gas, star formation and stellar feedback 
\citep[see, e.g.,][for discussions of this issue]{bower06,croton06,cattaneo07}.
In such `standard' models, star formation is insufficiently quenched, leading 
to a dramatic overabundance of massive blue star-forming galaxies.

Current models have attempted to remedy this shortcoming by quenching
star formation in galaxies through 
two main classes of mechanism.  The first mechanism, 
which affects only satellite
galaxies (i.e., the non-central galaxies in groups and clusters), is 
a shutoff of gas cooling in galaxies once they
fall into a larger halo.  Such a recipe has been in at least semi-analytic
models for some time \citep[see, e.g.,][]{cole00}
and produces a population of relatively low-mass highly-clustered 
red sequence galaxies.  These prescriptions are 
being currently tested using the SDSS and other 
surveys (e.g., \citealp{baldry06}, \citealp{weinmann06}, \citealp{haines07}), 
and I will not touch further on this issue in this paper. 

Another mechanism (or mechanisms) has been required
by the models to shut off star formation in central
galaxies (i.e., the galaxies lying in the center
of their dark matter halos, and presumably at the focus of 
any inflow of gas).  
An important candidate (I will touch upon others later) 
is feedback from accretion of matter
onto supermassive black holes, which 
either disrupts gas cooling in the galaxy
halo (e.g., `radio mode AGN feedback', \citealp{croton06}, \citealp{fabian06}) 
or drives gas out of 
the central galaxy (e.g., `quasar mode AGN feedback'; \citealp{kauf00}; 
\citealp{hopkins06}, \citealp{tremonti07})\footnote{It is expected
that such mechanisms act together; e.g., stellar or quasar-powered
winds may expel the gas from the galaxy originally, while
low-level AGN activity may suppress future gas cooling.  Most models
include both types of feedback. }.
In this picture, given 
the existence of a tight correlation between black hole mass
and bulge mass \citep[e.g.,][]{mag98,haering04}, it is natural to expect
that quenching and the existence of a large bulge would be
tightly linked.  

Indeed, such a correlation is well-documented, at least in a broad sense
(see, e.g., \citealp{strateva01} or Fig.\ 7 of \citealp{blanton03prop}).
Yet, if AGN feedback is the only way for central galaxies
to quench their star formation, such a paradigm makes a strong prediction: 
{\it bulgeless quenched central galaxies cannot exist.}  In this sense, 
the small but non-negligible population of bulgeless (low S\'ersic index) 
galaxies with red colors in Fig.\ 7 of \citet{blanton03prop} is 
of key importance.  Are these galaxies all satellite galaxies and/or 
dust-obscured edge-on galaxies?  Or, is there indeed a significant
population of red, central bulgeless disk galaxies?  
In the latter case, one would
be driven to prefer, at least in certain circumstances, 
other possible mechanisms for quenching star formation 
(see, e.g., \citealp{naab07}, \citealp{db07}, \citealp{khochfar07}, 
\citealp{dekel06}, \citealp{birnboim07} or
\citealp{guo07} on gravitational heating, the influence of the development 
of virial shocks, and the 
heating of large halos with cosmic ray energy).

\begin{figure*}[t]
\begin{center}
\epsfxsize 18.0cm
\epsfbox{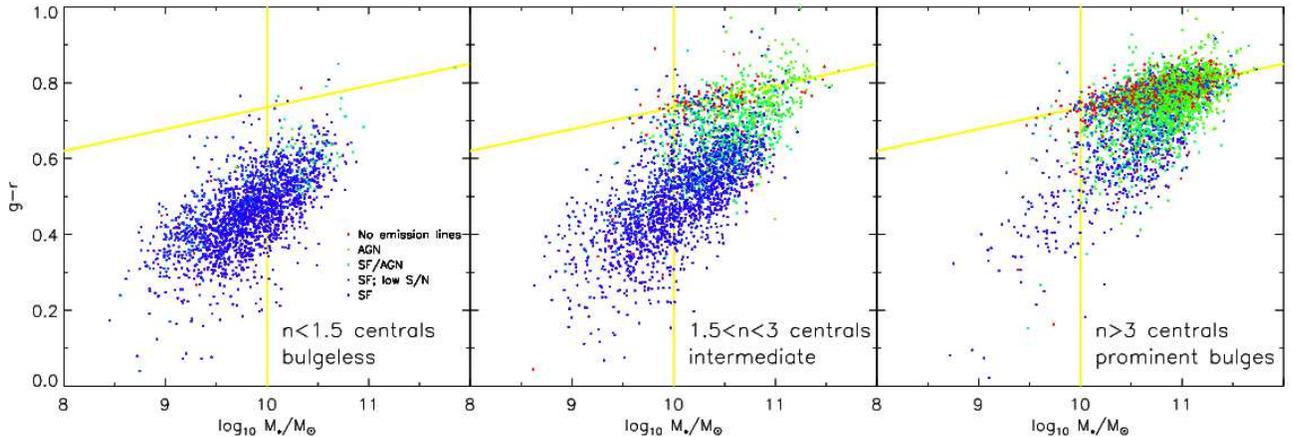}
\end{center}
\caption{\label{fig:colmass} 
The color--mass distribution for {\it central} galaxies with $b/a > 0.5$ and
$0.02<z<0.06$.  Galaxies are separated into those lacking
significant bulges (S\'ersic index $n<1.5$; left), those with 
intermediate bulge-to-disk ratio ($1.5 < n < 3$), and those 
with significant bulges ($n > 3$; right).  Colors
denote the emission-line classification classes
from \protect\citet{brinchmann04}.  Red denotes galaxies unclassified
because of a lack of line emission; the bulk of these galaxies are not
forming stars.  Hues of blue show galaxies with line emission characteristic
of being powered by star formation; green shades show star formation/AGN 
composites or AGN.  There are very few $n<1.5$ central red sequence galaxies, 
and almost all of these have an AGN.  It would appear that in order
to quench star formation completely, central galaxies must have 
a significant bulge and/or supermassive black hole.
}
\end{figure*}

\begin{figure*}[t]
\begin{center}
\includegraphics[width=3.5cm]{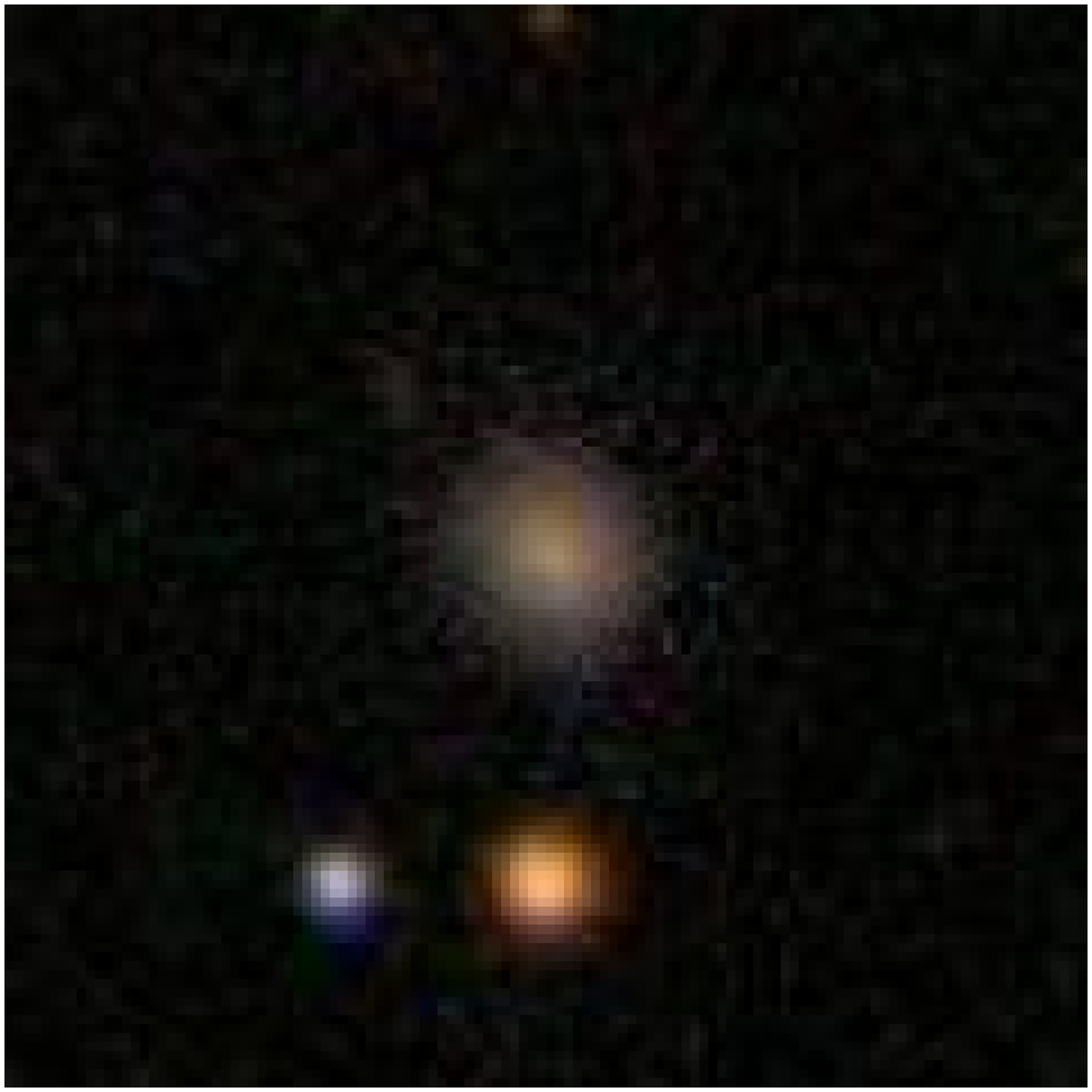}
\includegraphics[width=3.5cm]{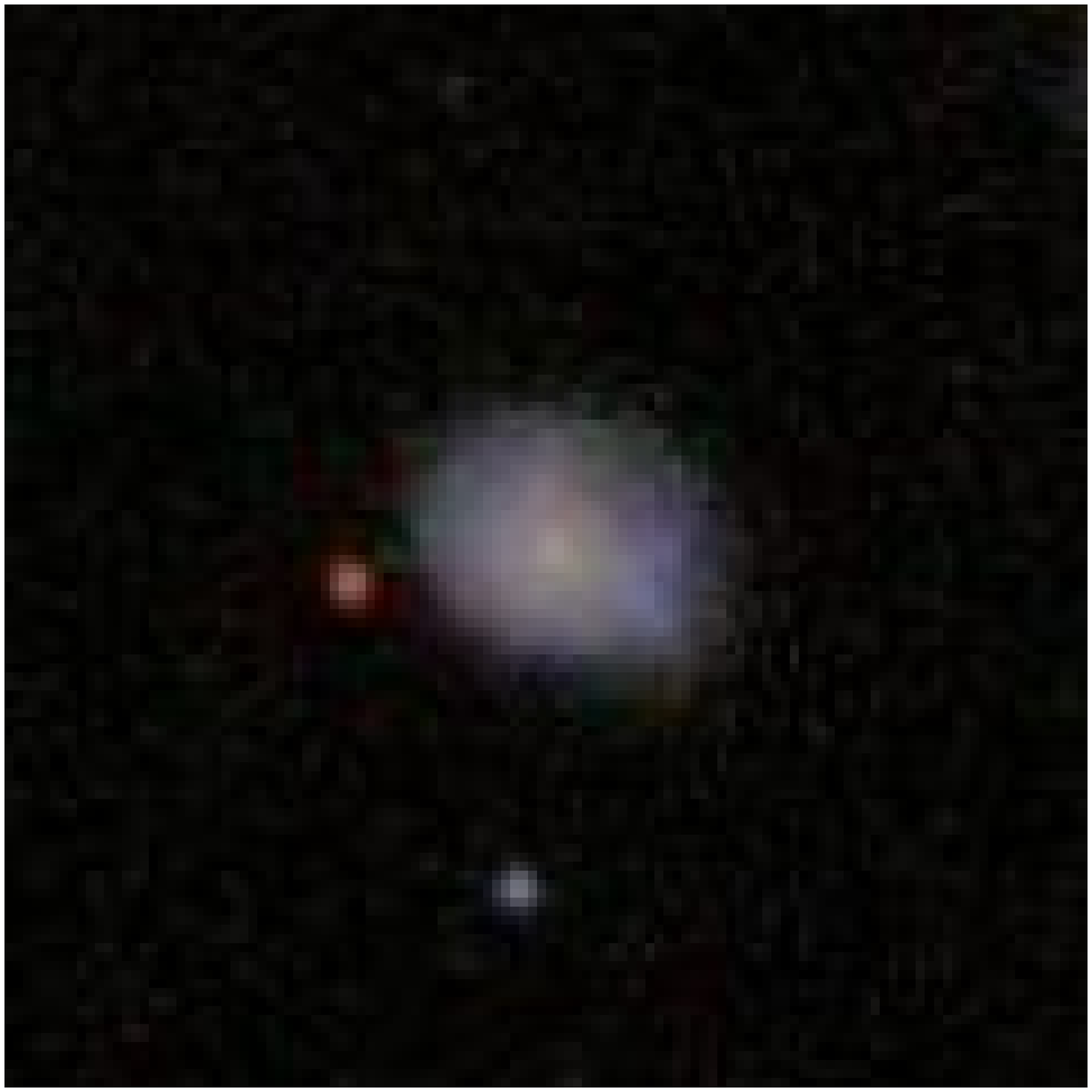}
\includegraphics[width=3.5cm]{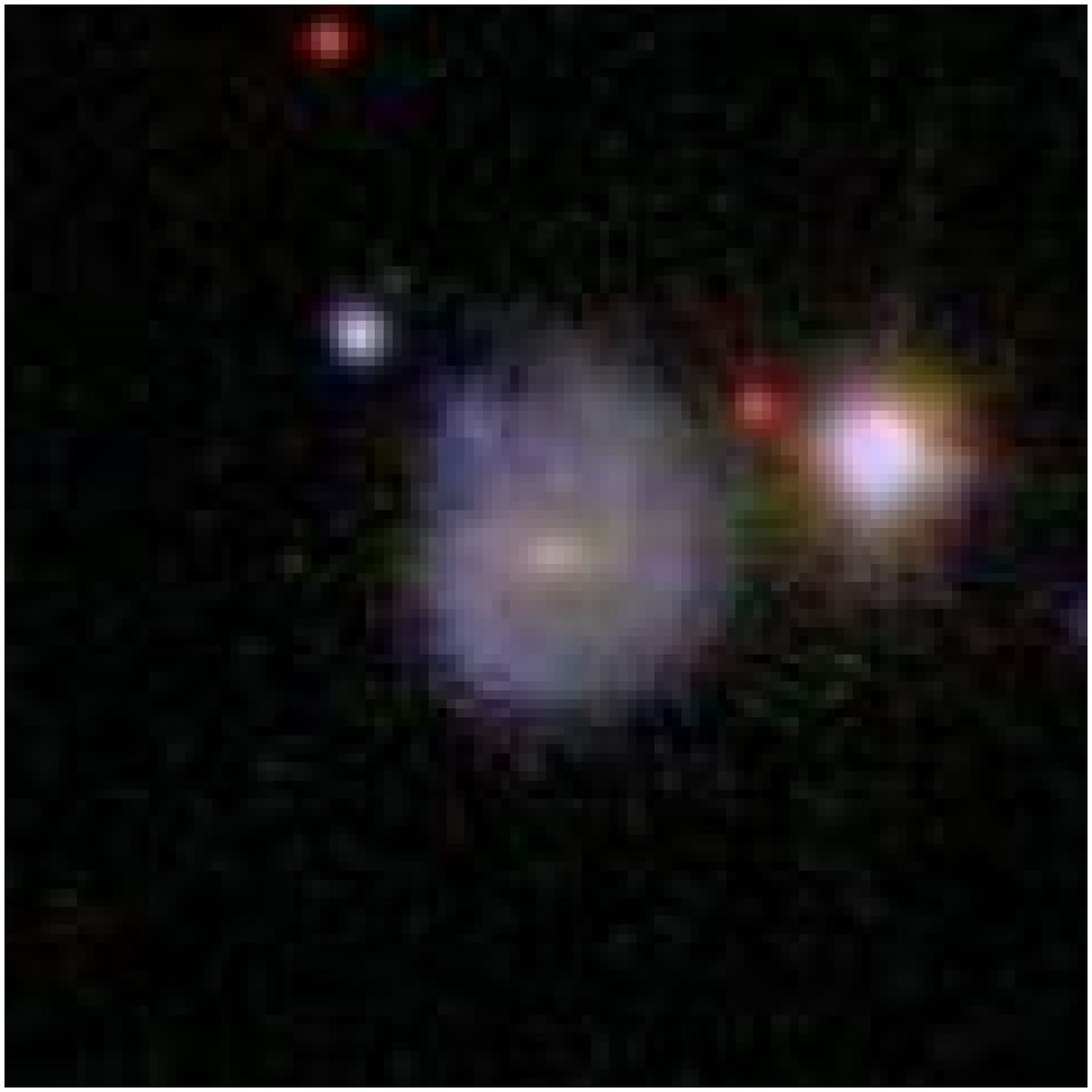}
\includegraphics[width=3.5cm]{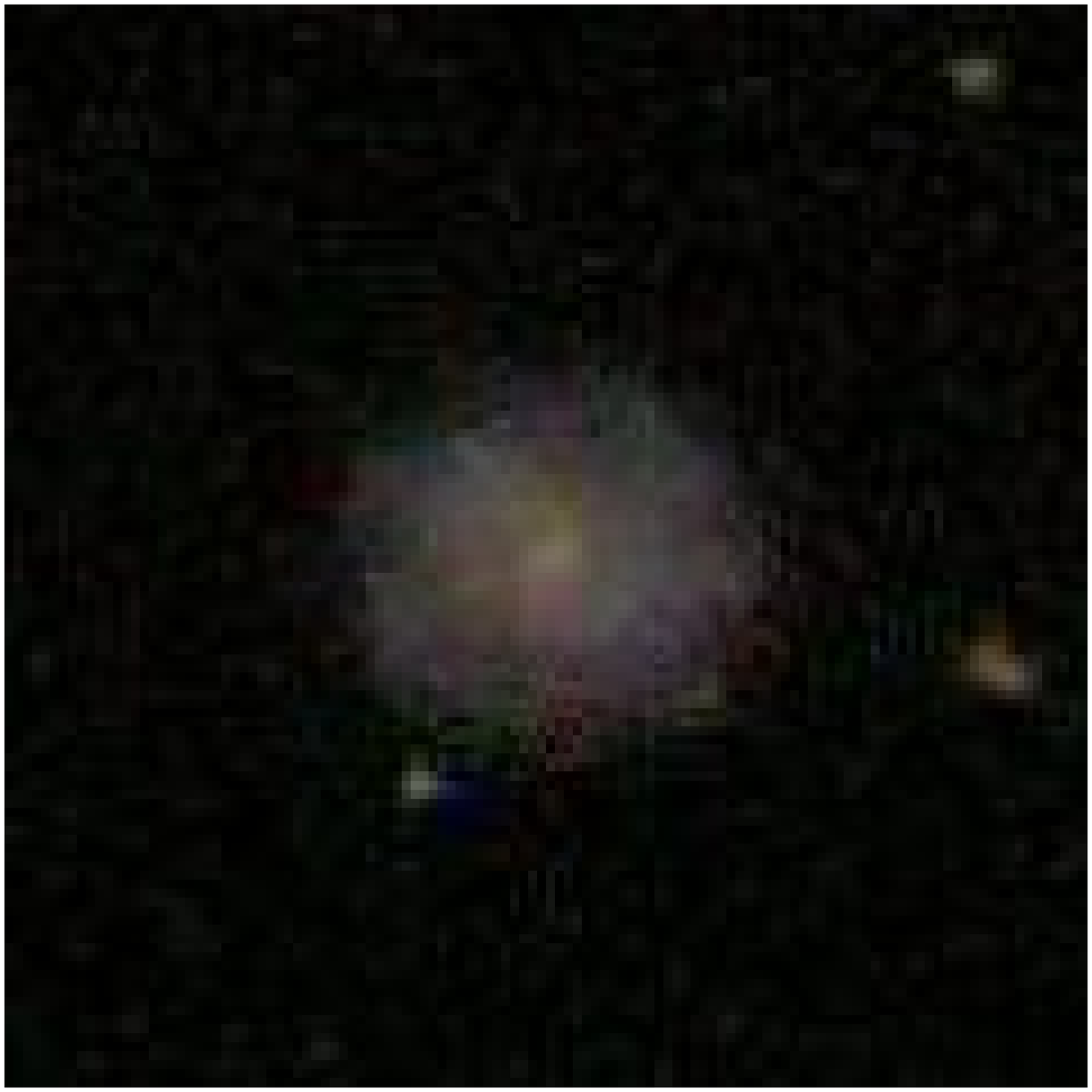}
\includegraphics[width=3.5cm]{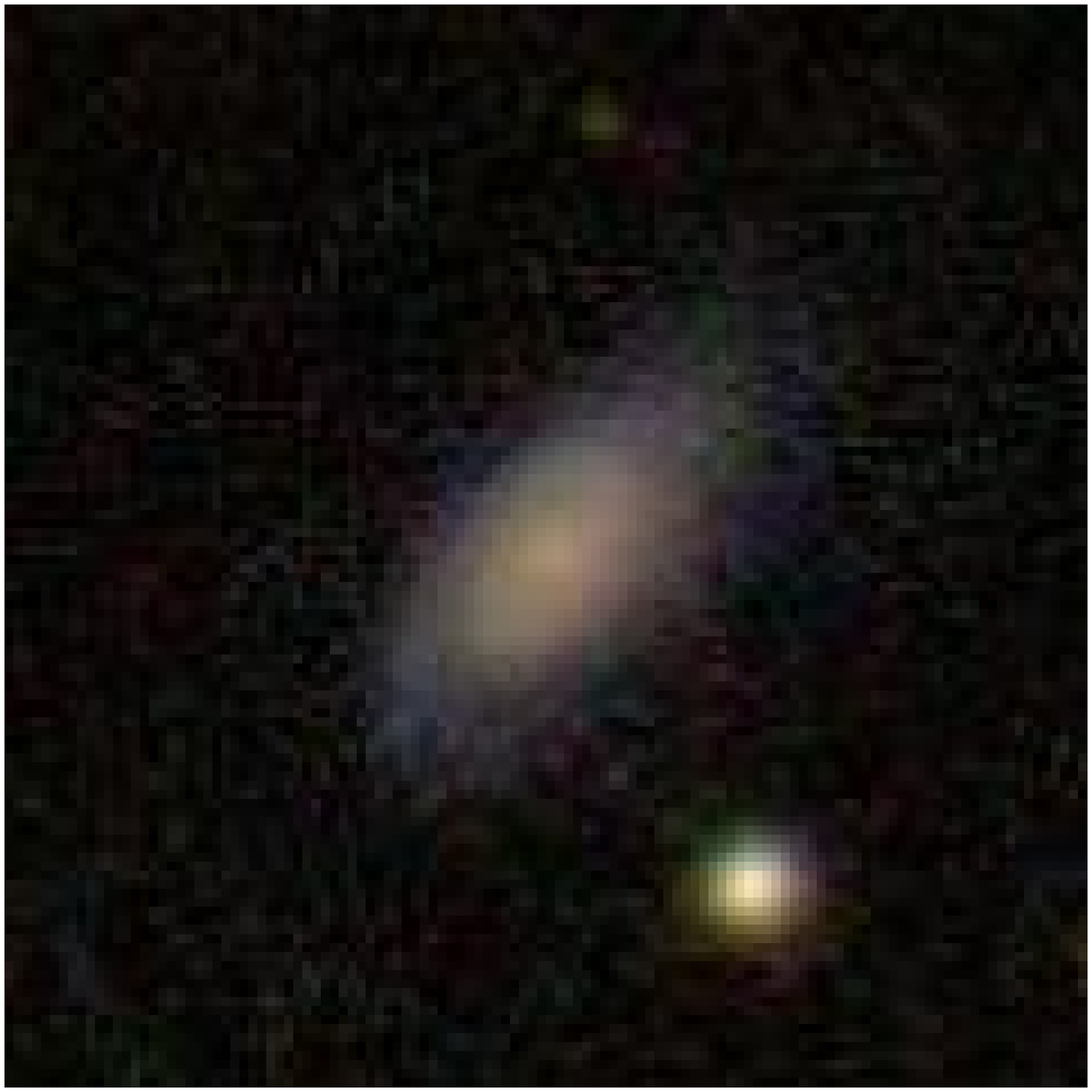}
\includegraphics[width=3.5cm]{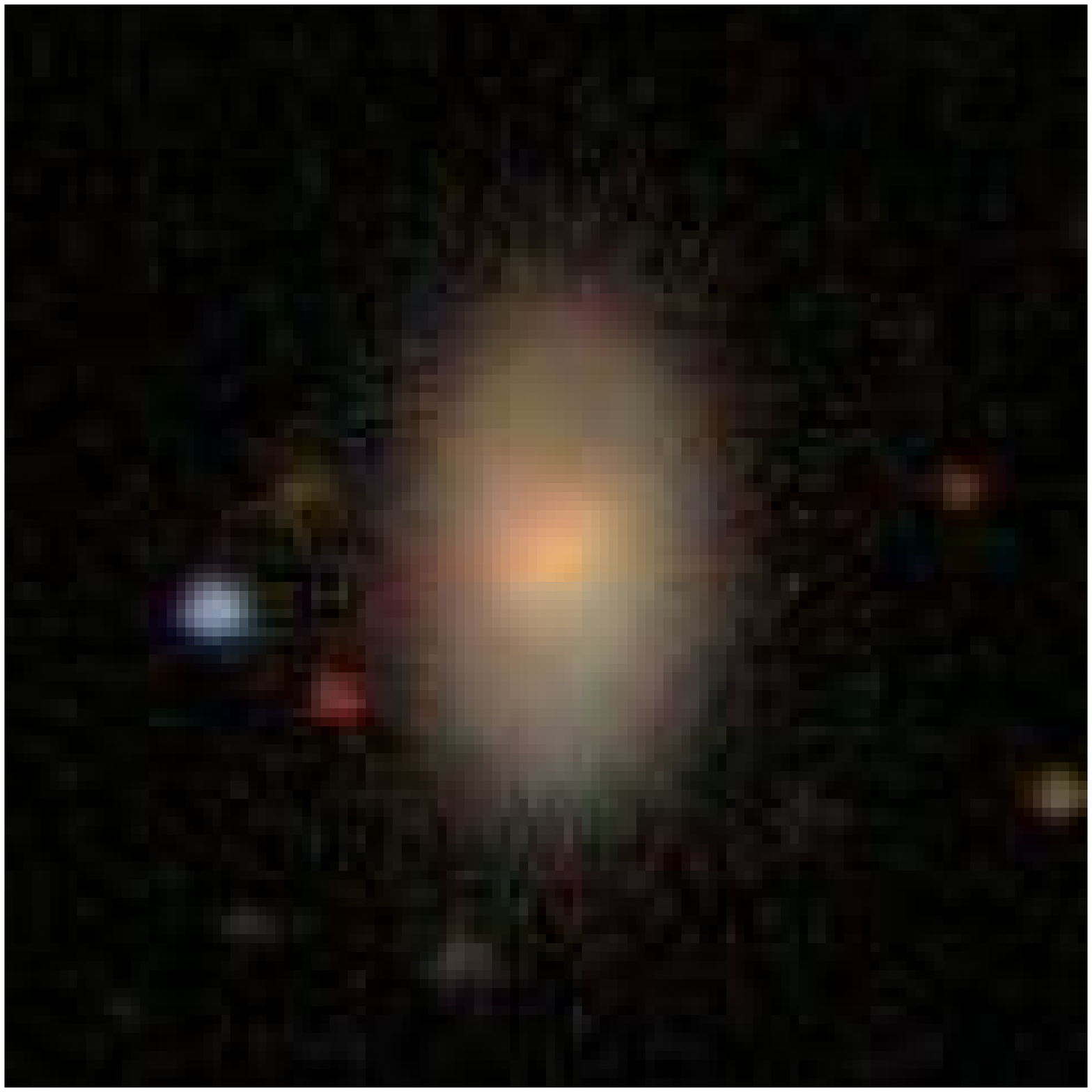}
\includegraphics[width=3.5cm]{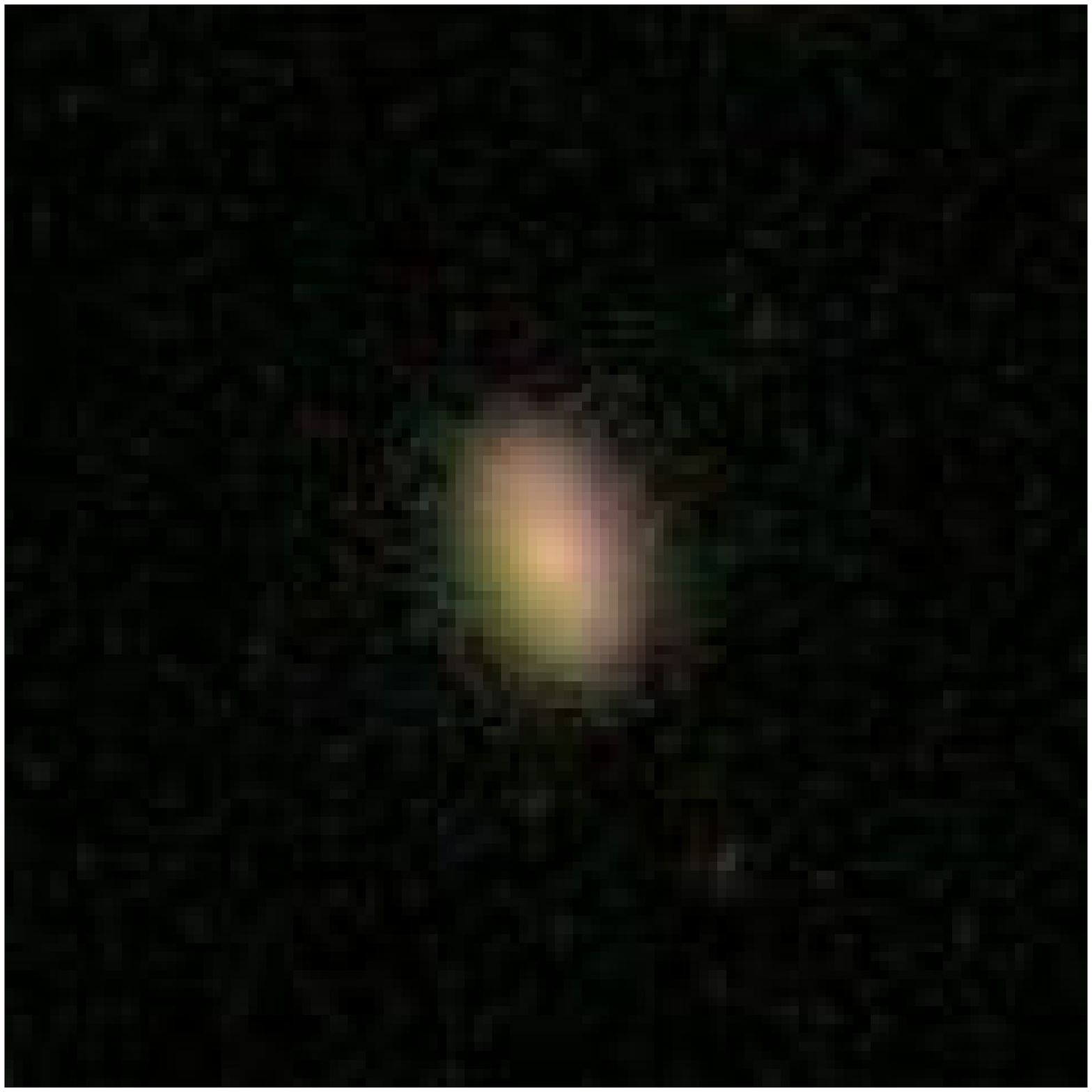}
\includegraphics[width=3.5cm]{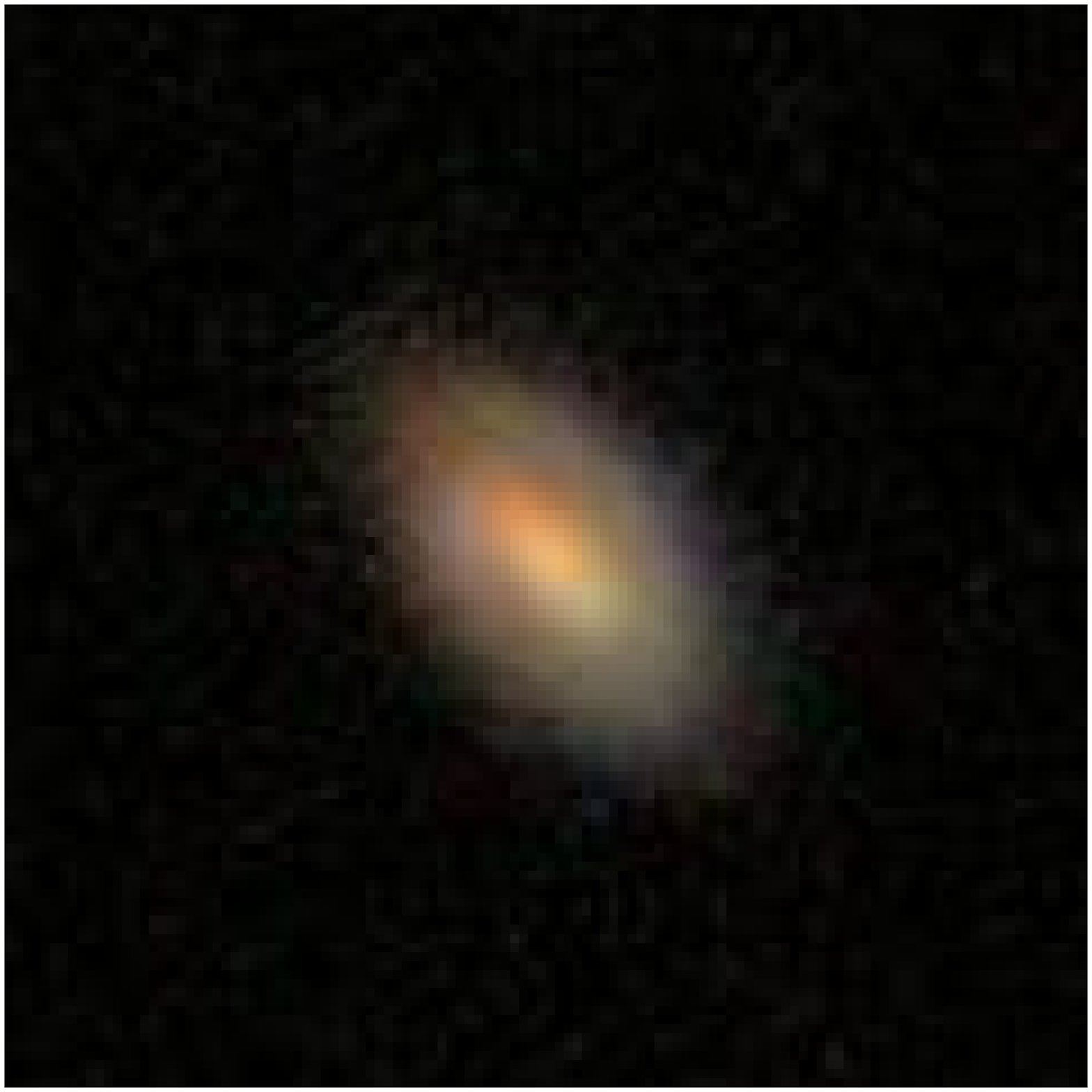}
\includegraphics[width=3.5cm]{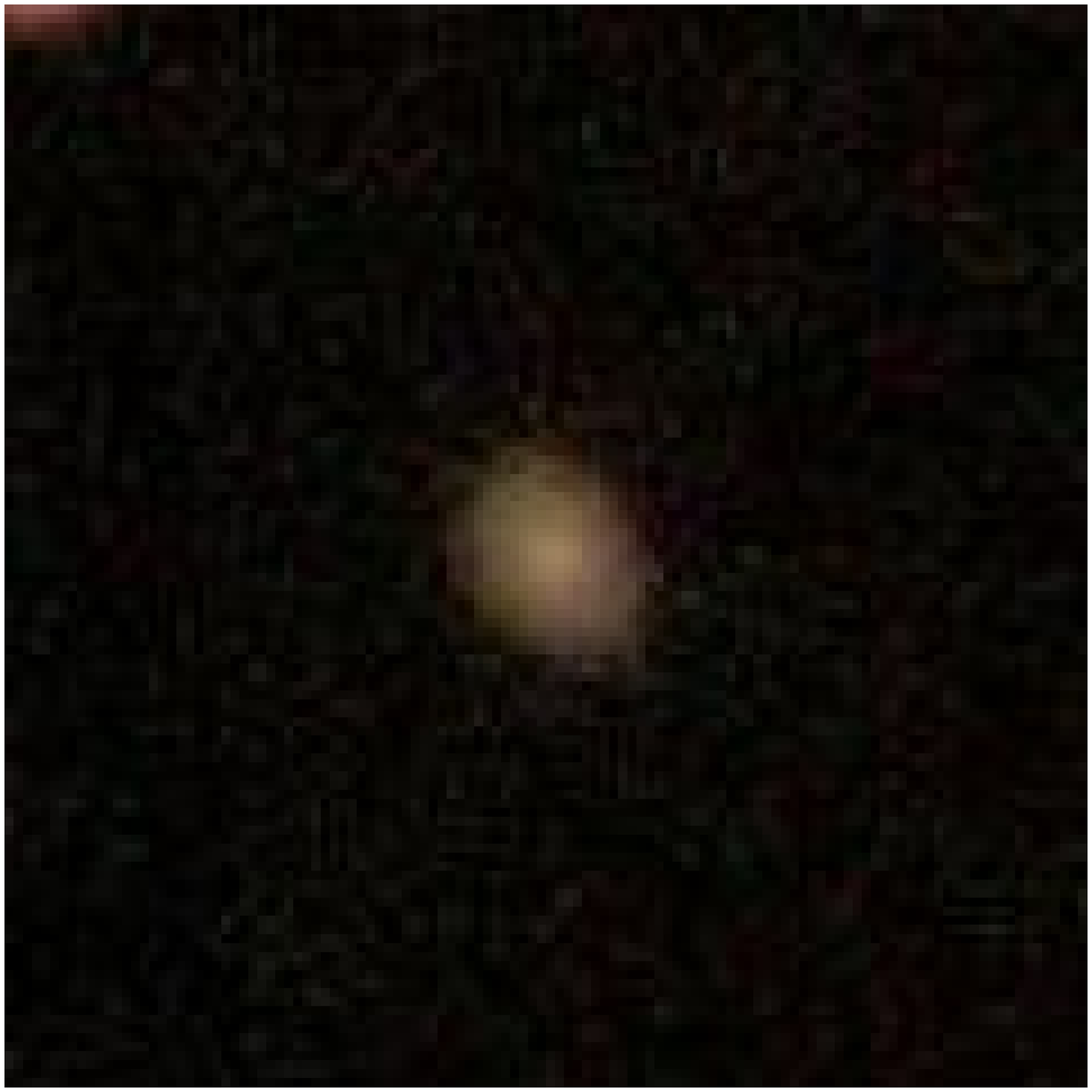}
\includegraphics[width=3.5cm]{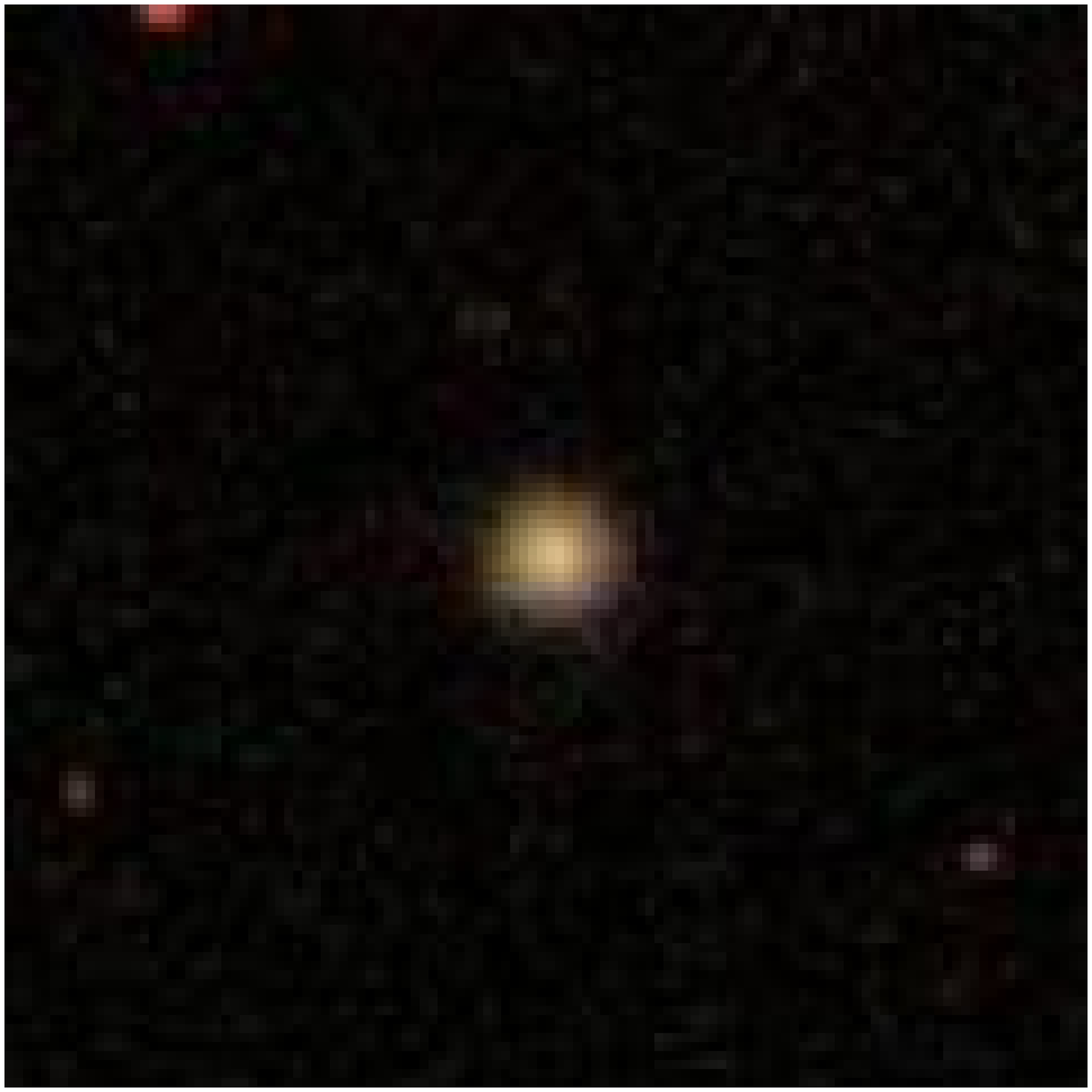}
\includegraphics[width=3.5cm]{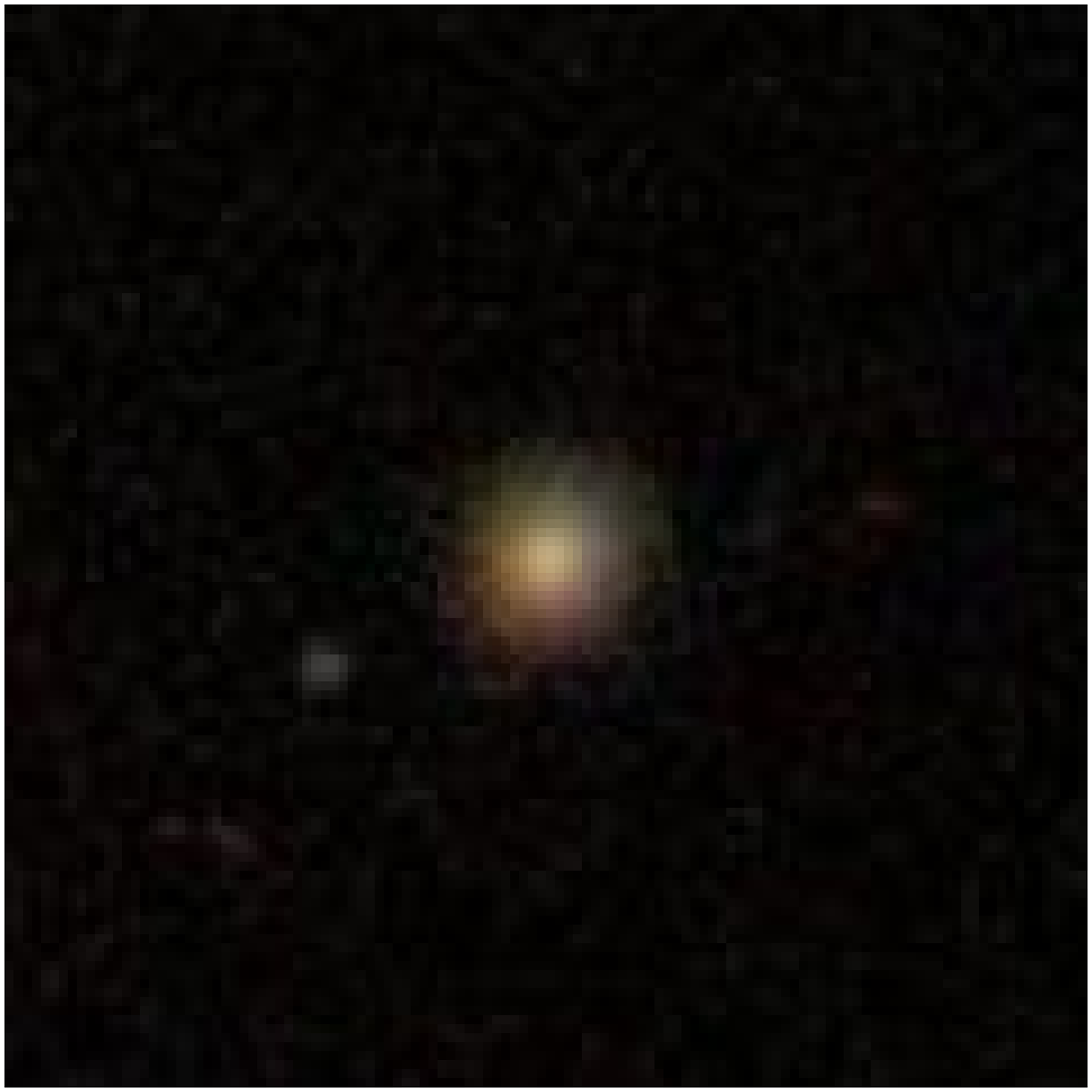}
\includegraphics[width=3.5cm]{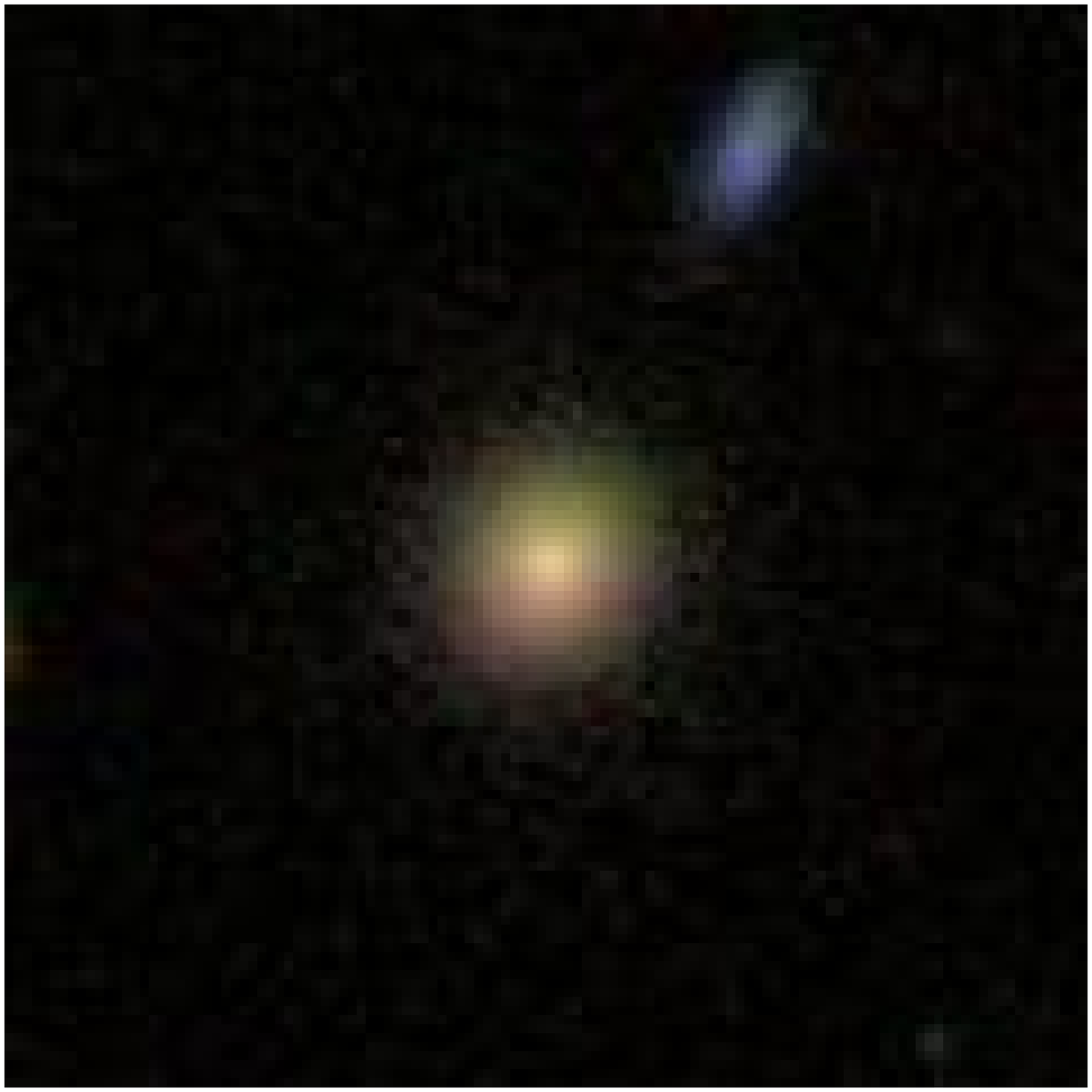}
\includegraphics[width=3.5cm]{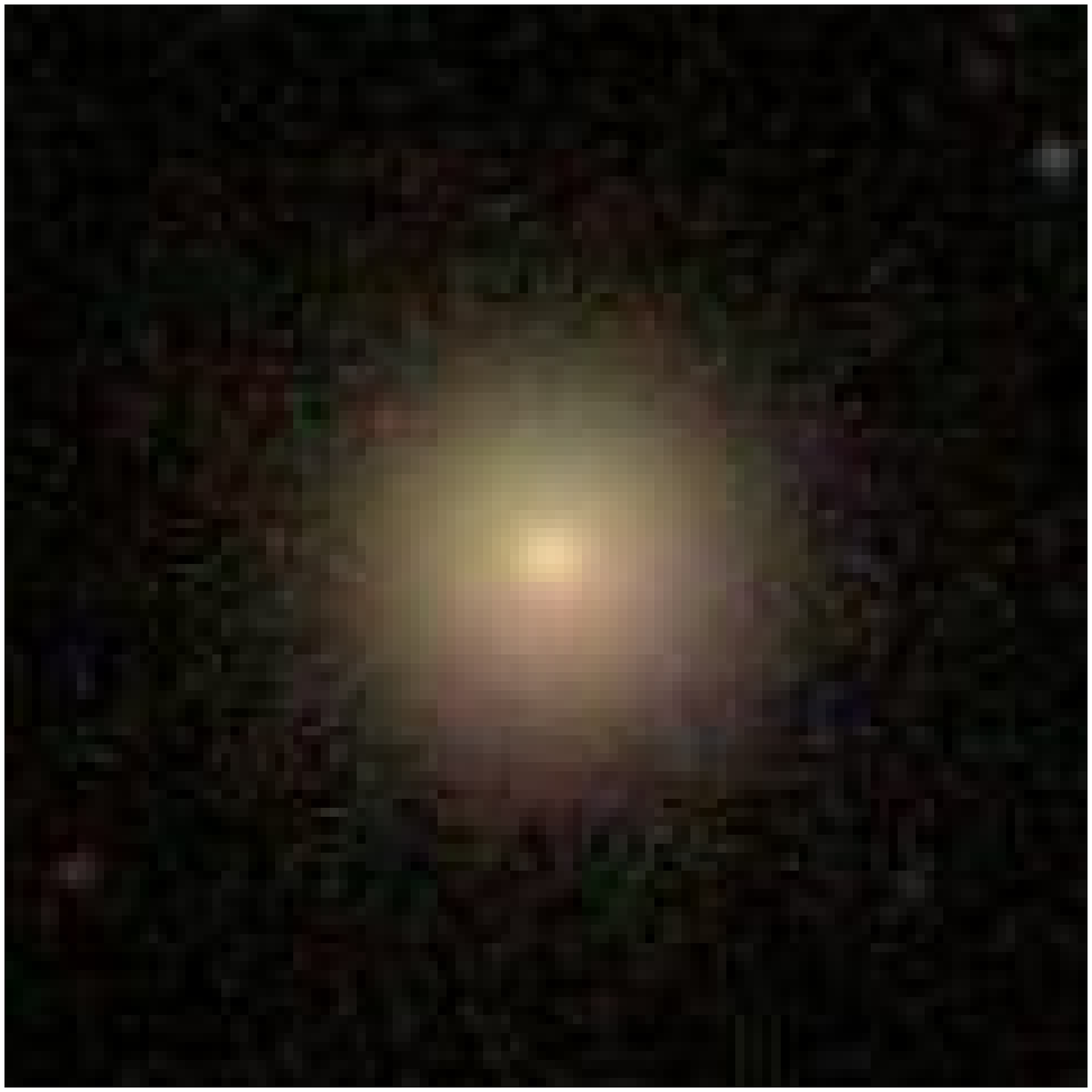}
\includegraphics[width=3.5cm]{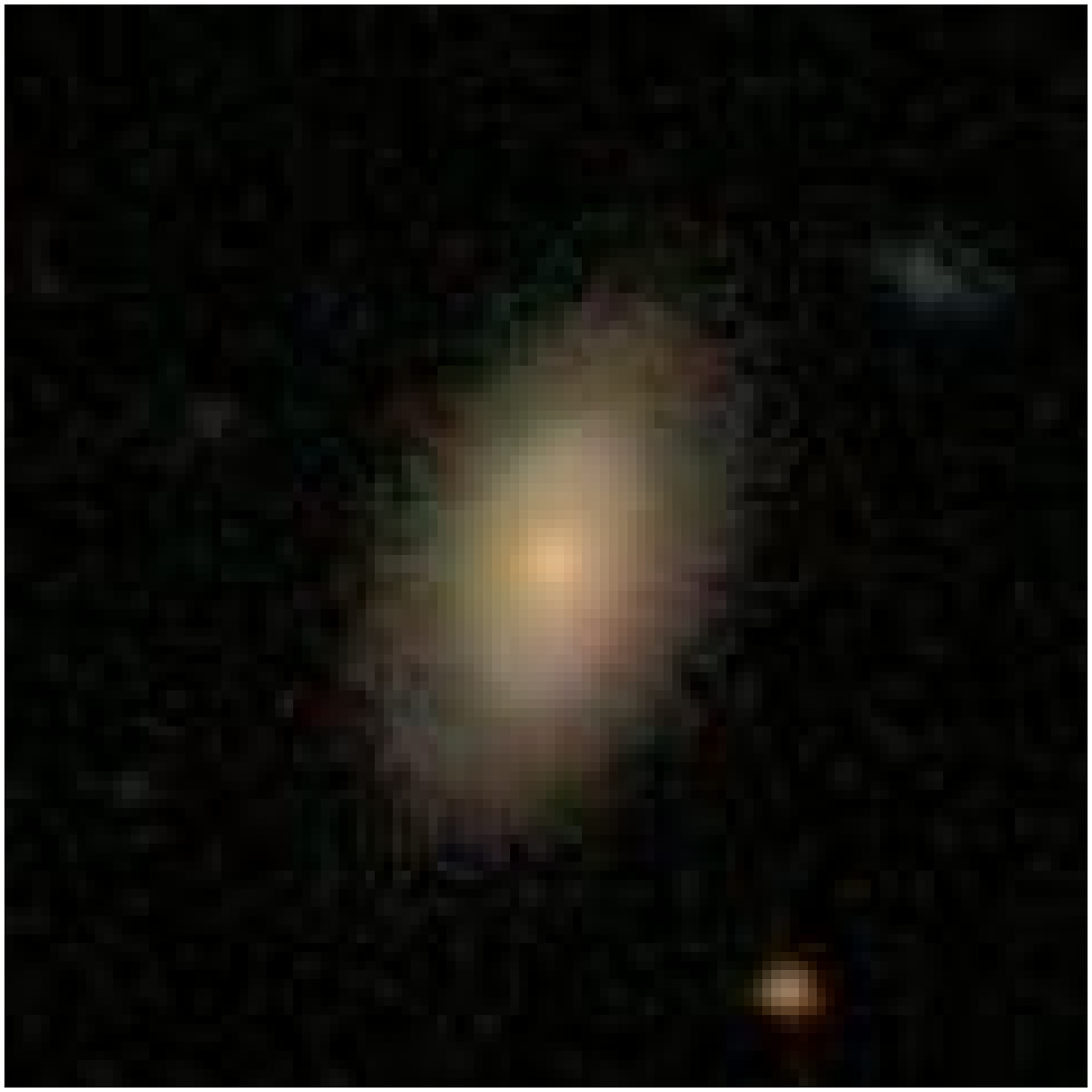}
\includegraphics[width=3.5cm]{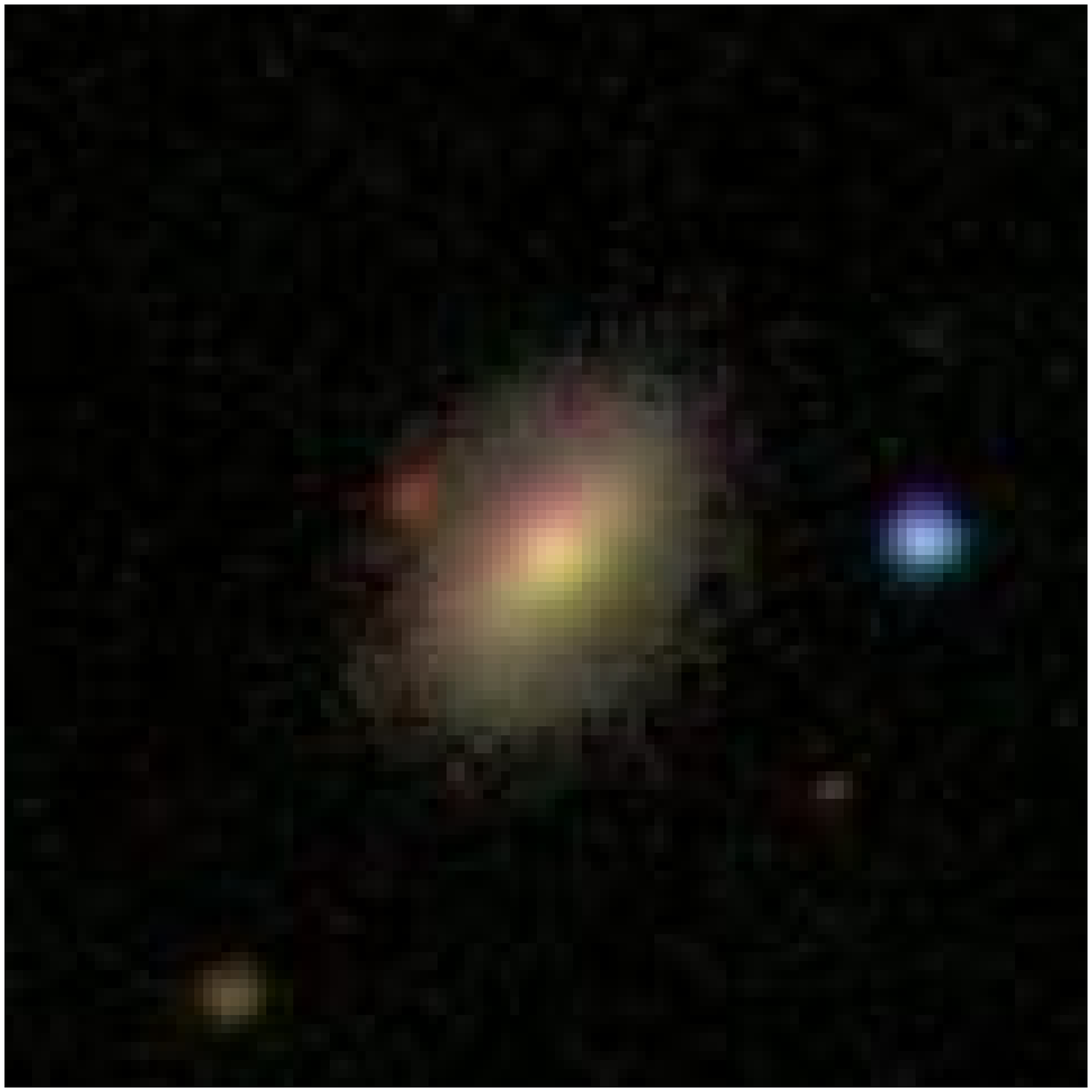}
\end{center}
\caption{\label{fig:pictures} 
SDSS cutouts of galaxies in the sample with $M_* > 10^{10} M_{\sun}$.
{\it Top:} Blue $n<1.5$ central galaxies.  {\it Middle:} Red $n<1.5$
central galaxies.  {\it Bottom:} Red $1.5\le n < 2.5$ central galaxies.
The rightmost galaxy in the middle row is the only red $n<1.5$
central galaxy lacking line emission.  Cutouts are roughly 40$"$ on a side.
There is a tendency for the red $n<1.5$ central galaxies to appear rather
more concentrated than their blue counterparts, lending weight to the 
notion that a significant bulge component is a requirement for 
quenching of star formation in central galaxies.  
}
\end{figure*}

The goal of this paper is to attempt to test
the key prediction of AGN feedback --- namely, that bulgeless quenched central 
galaxies do not exist --- using the Sloan Digital Sky Survey's 
Data Release 2 \citep{dr2}.
In \S \ref{sec:data}, I describe 
the data and derived parameters that I use in this investigation.
In \S \ref{sec:res}, I describe the results, showing the 
properties of the galaxy population as a function of galaxy 
morphology for central galaxies.  In sections  \ref{sec:disc} and 
\ref{sec:conc}, I discuss the results and present my conclusion.  
I adopt 
$H_{0}=70$\,km\,s$^{-1}$\,Mpc$^{-1}$,
$\Omega_\Lambda=0.7$ and $\Omega_{\rm m}=0.3$ in what follows.

\section{The Data} \label{sec:data}

I use publicly-available catalogs
derived from the Sloan Digital Sky Survey's (SDSS) Data Release
Two \citep{dr2}\footnote{DR2 is used in preference to 
larger, later SDSS data releases because it is the only
data release to have publicly-available galaxy density 
estimates, group catalogs, S\'ersic fits and star formation rate 
estimates.}.  The SDSS is an imaging and spectroscopic survey that has
so far mapped $\sim 1/4$ of the sky. Imaging data are produced simultaneously
in five photometric bands, namely $u$, $g$, $r$, $i$, and
$z$~\citep{fu96,gunn98,hogg01,gunn06}. The data are processed through
pipelines to measure photometric and astrometric properties
\citep{lupton99,st02,smith02,pier03,iv04,tucker06} and to select targets for
spectroscopic follow-up \citep{blanton03tiling,strauss02}.  
DR2 includes spectroscopy
over an area of $\sim 2600$ square degrees, and imaging for a larger
area. 

For this paper, I use the sample of galaxies in 
the New York University Value Added Galaxy Catalog (NYU VAGC;
\citealp{vagc}).  For galaxies with spectroscopic
redshifts, I use absolute magnitudes
$k$-corrected to the rest-frame $g$ and $r$ passbands \citep{kcorrect}
and corrected for Galactic foreground extinction following \citet{sfb98}.  
I adopt the $r$ absolute Petrosian magnitude for the
galaxy absolute magnitude (random and systematic 
uncertainties $\la 0.15$\,mag), and the $(g-r)$ model
color for a higher S/N estimate of galaxy color 
(uncertainties $\la 0.05$\,mag).  Stellar 
masses were estimated under the assumption 
of a universally-applicable \citet{chabrier} stellar IMF
 using the following
color--stellar M/L relation: 
\begin{equation}
\log_{10} {\rm M/L_r} = -0.406 + 1.097(g-r), 
\end{equation}
where an offset of $-0.1$ dex has been applied to the 
relation from \citet{bell03mf} to convert from a diet
Salpeter IMF to a \citet{chabrier} IMF. 
These stellar masses have random uncertainties of $\sim 0.1$\,dex; 
systematic uncertainties from recent bursts of star formation 
and other sources may exceed this estimate in cases where the
bursts contribute a significant fraction of the 
galaxy's light \citep{bdj}.

In order to 
gain a more complete understanding of star formation and AGN 
activity in the sample galaxies, star formation and AGN 
classifications and estimates of total star formation rate
were taken from \citet{brinchmann04}, using emission 
line measurements described in \citet{tremonti04}.  Galaxies are classified
as star forming, AGN, composites, or are left unclassified (typically
because the galaxies lack line emission in their SDSS spectra).

Disk galaxies were selected by selecting 
galaxies with nearly an exponential light profile.  \citet{blanton03prop}
fit the light profile of galaxies in the SDSS with a seeing-convolved
\citet{sersic} profile $\Sigma(r) \propto exp[-\kappa(\frac{r}{r_0}^{1/n}-1)]$. 
The S\'ersic index $n$ describes the shape of the light
profile, where $n=1$ corresponds to an exponential light profile
and $n=4$ corresponds to a $r^{1/4}$ law profile characteristic
of massive early-type galaxies.  In what follows, I apply a cut to 
the S\'ersic indices in $r$-band 
$n<1.5$ to select disk galaxies with little or no contribution 
from a bulge to the light profile of the galaxy; I apply a further
cut that the half-light radius should be larger than $0.4"$ to 
discard all galaxies with little/no structural information in the NYU 
VAGC\footnote{The S\'ersic fits are used in preference to the more
frequently used concentration parameter $c_r$ because the S\'ersic 
fits are convolved with the point spread function (i.e., the S\'ersic fits
should be less dependent on seeing than the seeing-dependent concentrations).}.

Finally, in order to test the model predictions for the properties
of central and satellite galaxies separately, I use a volume-limited
catalog of galaxy groups complete to
$\log_{10} M_{\rm group}/M_{\sun} > 11.7$ for group redshifts $z<0.06$
from \citet{yang05}.   
Groups are identified using an iterative method which adopts a trial
mass and identifies galaxies within a group mass-dependent transverse
radius and velocity difference. Then based on the total $r$-band luminosity
of all candidate group members the group mass is adjusted and 
the process repeated until one converges on a final group mass.  
The group mass is assigned based on the total $r$-band 
group luminosity of all members; in a cosmological simulation with the same
volume as the observed sample, the most luminous group is assigned
the largest halo mass, the second most luminous group the second
largest mass, etc.
Such groups are $\sim 90\%$ complete and contaminated at the 
$\sim 20\%$ level.  In this group catalog, the majority 
of galaxies reside in single galaxy `groups'; i.e., they
are the central galaxies in their own halos, with no satellites
above the luminosity limit of the sample. 
For the purposes of this paper, I am almost completely
unaffected by group mass uncertainties, as the primary purpose in using the 
catalogs is to separate the galaxy population into central 
galaxies and satellites; obviously, for single galaxy groups
the identification of the central galaxy is trivial. 
For all multi-galaxy
groups, Yang et al.\ identify the brightest galaxy in the group as the central
galaxy.  Comparison with mock group catalogs derived using this
method demonstrates that $>97.5\%$ of central galaxies are indeed
the brightest galaxy in their group \citep{weinmann06}.  
Satellites are defined to be 
all other galaxies in groups.

\section{Results} \label{sec:res}

The goal of this paper is to test the generic prediction
of AGN feedback that there should be no 
bulgeless central galaxies with quenched star
formation.  In order to test this prediction, I select galaxies 
with $0.02<z<0.06$ from the above sample.  I further select 
galaxies with $\la 60\arcdeg$ inclination ($b/a > 0.5$).
This is a selection cut of some importance, as it minimizes
the effects of dust extinction on galaxy colors and 
emission line-derived estimates of star formation rates; 
low-inclination galaxies
with red colors and/or low star formation rates
become then a reasonably clean probe of quenching\footnote{A further
advantage of the selection of low-inclination systems
is that the circular apertures used for the SDSS model and S\'ersic fits
are more appropriate for these more 
circular systems than they are for higher inclination systems.}.  

The key result of this study 
is shown in Fig.\ \ref{fig:colmass}.  
 The left-hand panel shows the 
properties of central galaxies without
significant bulges, selected by requiring
that the S\'ersic index $n<1.5$\footnote{The referee
pointed out that there are a number of examples
of low-mass red sequence early-type (E/S0) galaxies with $n<1.5$, 
e.g., in the Virgo cluster.  Indeed, in the current sample, 
the vast majority of the red, non-star-forming $n<1.5$ 
galaxies are low-mass ($M_* < 10^{10} M_{\sun}$)
satellites, primarily of groups with group masses in excess of 
$10^{13} M_{\sun}$.  There are no $n<1.5$ non-star-forming 
central galaxies with $M_* < 10^{10} M_{\sun}$ (Fig.\ 
\ref{fig:colmass}, left panel), lending weight to the notion that 
while non-star-forming $n<1.5$ E/S0 galaxies exist, they 
exist almost exclusively as satellites of reasonably 
massive groups, and are the product of (primarily
hydrodynamical) stripping of late-type disk-dominated galaxies.
}.  
The right-hand panel shows the 
properties of central galaxies with large bulges, selected using
$n>3$.  It is to be noted that the $n>3$ bin contains a number of
galaxies with prominent (even dominant) disks; the $n>3$ is 
reflecting primarily the existence of a large 
bulge.  The middle panel shows the intermediate $1.5 < n < 3$ bin.
One can see that {\it there are very
few bulgeless galaxies on the red sequence}; adopting 
a definition of red sequence of $(g-r) > 0.57+0.0575\log_{10}(M_*/10^8M_{\sun})$
(i.e., a line 0.05\,mag
bluewards of the red sequence locus described by the thick
line) and restricting our attention only to galaxies with 
$\log_{10} M_*/M_{\sun} > 10$, only $0.5\%$ (14/2744) of red sequence central 
galaxies have $n < 1.5$ (14/725, or 1.9\%, of $n<1.5$ galaxies 
are on the red sequence)\footnote{There are 16 red sequence
central galaxies with values of $b/a > 0.5$ and $n<1.5$; two 
of these (SDSS J142922.73$-$004939.1 and SDSS J084958.78$+$381203.2) are
clearly edge-on and should have not been included in the sample (their
catalog values of $b/a$ are simply incorrect) and they were excluded
from the sample.}.  

Fig.\ \ref{fig:pictures} shows example $gri$ SDSS cutouts for subsamples
of central galaxies with $M_* > 10^{10} M_{\sun}$.  The top panels
show five of the 711 blue central $n<1.5$ galaxies; one can clearly
see that the typical galaxy in this subsample is a late-type disk
galaxy with a small or nonexistent bulge.  The middle panels
show five of the 14 red central $n<1.5$ galaxies.  One can see in
some cases a prominent bar and possible bulge.  The lower panels
show five of the 163 red central galaxies with $1.5 \le n < 2.5$.
There is a tendency towards better defined bulges than for the 
$n<1.5$ red central galaxy subsample; however, it is clear that at
least some fraction of the low-$n$ red subsample have structures
similar to the higher $n$ subsample, lending weight to the notion
that some of the 14 $n<1.5$ red central galaxies have 
incorrect $n$ estimates --- 0.5\% is an {\it upper limit}
to the fraction of bulgeless red sequence galaxies.  Taking 
Figs.\ \ref{fig:colmass} and \ref{fig:pictures} together, it
is clear that there are very few, if any, bona fide bulgeless
red central galaxies --- in agreement with the most basic 
expectation of the AGN feedback paradigm.

One may have the concern that perhaps this result was an artifact
of the way in which central and satellite galaxies are chosen by 
the group finding algorithm.  One can repeat the analysis 
using cylindrical overdensities from the NYU VAGC (i.e., overdensities measured
in cylinders of 1.4\,Mpc radius and 1600\,kms$^{-1}$ extent along the line 
of sight; \citealp{blanton03prop}), which are simpler and 
more conventional measures of environment.  
Isolated galaxies (defined
as having overdensities $\delta < 0$) are likely to reside as centrals
in their own halos.  When repeating the analysis using isolated
galaxies as a (conceptually less optimal) proxy for central 
galaxies, one finds that the fraction of $M_* > 10^{10} M_{\sun}$
red sequence galaxies in isolated environments with $n<1.5$ is 
1.0\% (18/1842).  So one finds, using a completely different technique
for identifying likely central galaxies, the same result: in central
galaxies, quenching is empirically correlated with the existence
of a prominent bulge component.

\section{Discussion}
\label{sec:disc}

\begin{figure}[t]
\begin{center}
\plotone{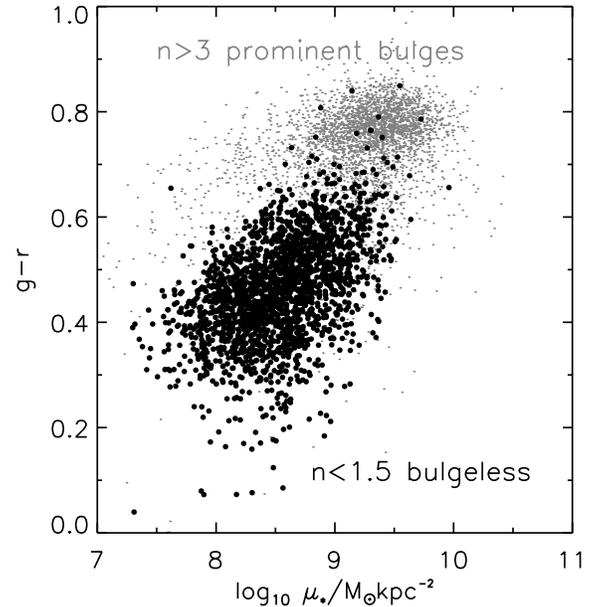}
\end{center}
\caption{\label{fig:coldens} 
The color--surface density distribution for 
central galaxies with $b/a > 0.5$ and
$0.02<z<0.06$.  Galaxies with large bulges ($n>3$) are shown in 
gray, and bulgeless galaxies ($n<1.5$) are shown in black.
Quenching is not driven by surface density; at all surface densities, galaxies
with prominent bulges $n>3$ have much redder colors than $n<1.5$ systems.
}
\end{figure}

\begin{figure}[t]
\begin{center}
\plotone{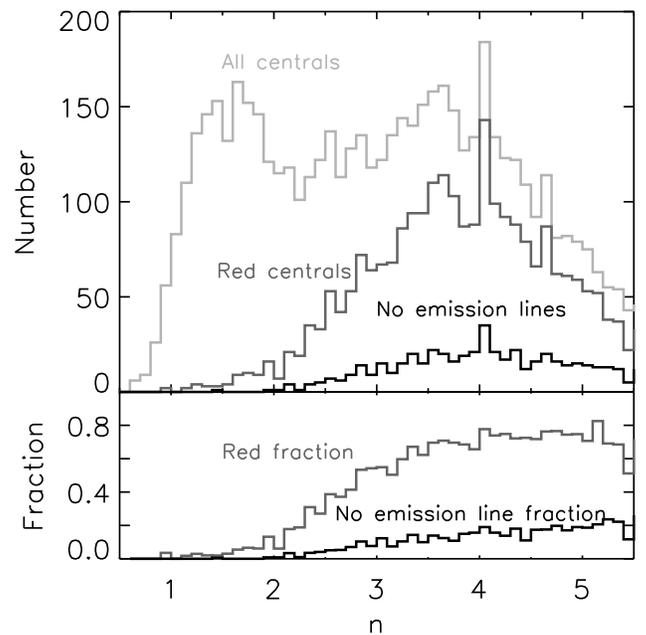}
\end{center}
\caption{\label{fig:nhist} 
The upper panel shows the 
histogram of S\'ersic values for all central galaxies
(gray), red central galaxies (dark gray), and 
galaxies with no detected emission lines (black).  
In the lower panel, the fraction of central galaxies 
that is red (dark gray) and that have no emission lines
(black) is shown as a function of S\'ersic index.
}
\end{figure}

It is worth asking if this empirical association between having a bulge and 
the ability to quench is a truly bulge-specific association, or if 
it reflects a third, unexplored parameter.  An obvious candidate
parameter is surface density.  In \citet{kauf03_dens} and \citet{BelldeJ00}
it was found that the stellar surface density of a galaxy was
correlated strongly with star formation history, inasmuch as
galaxies with high surface densities appeared to form the
bulk of their stars at earlier epochs than galaxies with
lower surface densities.  In Fig.\ \ref{fig:coldens}, 
I plot optical $g-r$ rest-frame color as a function 
of the stellar surface density $\mu_*$ within the half
light radius for bulgeless galaxies ($n<1.5$; black) and 
galaxies with large bulges ($n>3$; gray).  While $g-r$
is indeed a function of surface density for the $n<1.5$ galaxies, 
it is clear that at a given surface density (e.g., in the range around
$\sim 10^9 {\rm M_{\sun}kpc^{-2}}$) that galaxies with strong
bulges are primarily quenched, whereas bulgeless galaxies
at that same density still actively form stars.  Such a conclusion
was also reached by \citet[see their Fig.\ 14]{kauffmann06}, who explored
this star formation rate per unit stellar mass as a function of 
surface density for a samples of galaxy split by concentration.
Thus, one can rule out that the correlation seen in Fig.\
\ref{fig:colmass} is driven by underlying correlations with
 stellar surface density.

Fig.\ \ref{fig:nhist} shows the distribution of the S\'ersic indices
of the sample of central galaxies.  One can see that the fraction 
of the sample that is red is a strong function of the S\'ersic index, 
with a small red fraction for $n<2$ smoothly increasing to 
$n \sim 4$\footnote{The red fraction does not asymptotically approach unity 
for the high $n$ systems because of 
the combined influence of our somewhat restrictive red sequence cut 
and the scatter of a small fraction of high $n$ systems towards the blue, 
as seen in Fig.\ \ref{fig:colmass}.}.  Interestingly, 
only a small fraction ($\sim 1/4$) of red galaxies
have no emission lines at all (the fraction of galaxies which are
completely quenched, at least to the detection limits characteristic of 
the SDSS; see \citealp{haines07rs} for the same result
using GALEX data in conjunction with the SDSS).  
Put differently, complete quenching of gas infall into  
in central galaxies is a relatively uncommon occurrence; most red sequence
centrals have some modest AGN activity or low-level star formation.  
In particular, only 1 out of the 14 red central $n<1.5$ galaxies
with stellar masses above $10^{10} M_{\sun}$ lacks 
line emission Fig.\ \ref{fig:colmass}; i.e., complete quenching of bulgeless 
galaxies is a very rare occurrence\footnote{The right-most panel 
of the middle row of Fig.\ \ref{fig:pictures} shows this galaxy; this
galaxy appears to be significantly more compact and concentrated
than its $n<1.5$ would imply.}.  

Finally, it is obvious that while bulges appear to be a {\it necessary}
condition for the suppression of star formation, the simple
existence of a bulge is not a {\it sufficient} 
condition to suppress star formation.
There are a number of suggestions as to what this 
`second parameter' could be: pseudobulges vs.\ classical
bulges \citep{drory07}, the transition to a hot virialized halo
at halo masses $\sim 10^{12} M_{\sun}$ \citep{dekel06,cattaneo06},
or the rate at which cosmological infall can deposit energy into 
the halo \citep{naab07,khochfar07}.   In this context, 
it is noteworthy that for group masses $> 10^{13} M_{\sun}$ 23\% of 
the $n>3$ central galaxies are blue\footnote{Most of these
blue $n>3$ central galaxies are spiral galaxies with large bulges (70\%), 20\%
appear to be relatively undisturbed early-type (E/S0) galaxies, and 10\%
of them are highly disturbed, with asymmetries or tidal tails indicative
of a previous interaction or merger.} (as opposed to 35\% for 
halos with masses below $10^{13} M_{\sun}$).  Thus although 
models which assume quenching is complete above 
a given mass cut (e.g.,  $10^{12} M_{\sun}$ in the work of 
\citealp{cattaneo06}) offer a qualitatively interesting
and computationally convenient approximation for the evolution 
of the galaxy population, it is clear that such an approximation 
is not correct in detail --- even at group masses $>10^{13} M_{\sun}$
quenching is incomplete.

\section{Concluding Remarks} \label{sec:conc}

The main message of this paper is simple.  Galaxy evolution models 
with AGN feedback predict that quenching of star formation in central
galaxies should occur only in those central galaxies with 
prominent bulges (therefore, large supermassive black holes).
I have shown that this prediction was essentially correct, using
data from the SDSS DR2.
I choose to analyze only relatively low-inclination galaxies with $b/a>0.5$; 
while the general conclusions of this paper would hold also 
if higher inclination galaxies were included, the scattering
of $n<1.5$ galaxies onto the red sequence by dust extinction 
considerably dilutes the significance and impact of the results.
I found that at least $99.5\%$ of red sequence 
galaxies with stellar masses $>10^{10} M_{\sun}$ 
have prominent bulges (S\'ersic indices $n>1.5$).
Furthermore, most the 0.5\% of 
bulgeless red sequence central galaxies had 
detectable line emission; 
there is only 1 $n<1.5$ central galaxy without detectable star formation 
out of the 6036 central galaxies in this mass-limited sample.  This almost
perfect empirical 
association between having a prominent bulge and quenching is in 
excellent agreement with the basic expectation of AGN 
feedback.

\acknowledgements
I thank the referee for their constructive comments and questions.
I wish to particularly thank Julianne Dalcanton
for long and fruitful discussions and comments on early drafts
of this paper; I wish to acknowledge also useful discussions with 
Frank van den Bosch, 
Eva Schinnerer, Hans-Walter Rix, David Hogg and 
Darren Croton.  I also wish to extend my deep
appreciation to 
Michael Blanton, Jarle Brinchmann, Xiaohu Yang, 
and the teams which produced and 
tested the value-added
public data products which made this work possible.  
This work was supported by the Emmy Noether Programme of the Deutsche
Forschungsgemeinschaft.  

Funding for the Sloan Digital Sky Survey (SDSS) has been provided by the Alfred P. Sloan Foundation, the Participating Institutions, the National Aeronautics and Space Administration, the National Science Foundation, the U.S. Department of Energy, the Japanese Monbukagakusho, and the Max Planck Society. The SDSS Web site is http://www.sdss.org/.

The SDSS is managed by the Astrophysical Research Consortium (ARC) for the Participating Institutions. The Participating Institutions are The University of Chicago, Fermilab, the Institute for Advanced Study, the Japan Participation Group, The Johns Hopkins University, Los Alamos National Laboratory, the Max-Planck-Institute for Astronomy (MPIA), the Max-Planck-Institute for Astrophysics (MPA), New Mexico State University, University of Pittsburgh, Princeton University, the United States Naval Observatory, and the University of Washington.

\end{document}